\makeatletter
\g@addto@macro\normalsize{%
  \setlength\abovedisplayskip{5pt}
  \setlength\belowdisplayskip{5pt}
  \setlength\abovedisplayshortskip{1pt}
  \setlength\belowdisplayshortskip{1pt}
} \makeatother

\documentclass[useAMS,referee,usenatbib]{biom}
\usepackage{graphicx}
\graphicspath{{./figs/}} 
\usepackage{float,subfigure}
\usepackage{mathrsfs,amsmath}
\usepackage{amssymb} 
\usepackage{gensymb} 
\usepackage{multirow}
\usepackage{etoolbox}
\usepackage{prodint}
\usepackage{color,soul}

\BeforeBeginEnvironment{figure}{\vskip-2ex}
\AfterEndEnvironment{figure}{\vskip-2ex}

\makeatletter
\g@addto@macro\normalsize{%
  \setlength\abovedisplayskip{3pt}
  \setlength\belowdisplayskip{3pt}
  \setlength\abovedisplayshortskip{2pt}
  \setlength\belowdisplayshortskip{2pt}
} \makeatother
\makeatletter

\BeforeBeginEnvironment{figure}{\vskip-2ex}
\AfterEndEnvironment{figure}{\vskip-2ex}

\newcommand{\nc}{\newcommand}
\nc{\nn}{\nonumber} \nc{\fns}{\footnotesize}

\nc{\revisionline}{\vspace{.1in} \today \vspace{.1in}
\hrule\hrule\hrule\vspace{.1in}} \nc{\newpp}{\vspace{.1in}
\noindent}

\nc{\slideline}{\smallskip \hrule\hrule \smallskip}

\nc{\wh}{\widehat} \nc{\wt}{\widetilde}

\nc{\bgam}{\pmb{\gamma}} \nc{\bgamma}{\pmb{\gamma}}
\nc{\bGamma}{\pmb{\Gamma}}

 \nc{\beq}{\begin{eqnarray*}}
\nc{\eeq}{\end{eqnarray*}}

\nc{\beqna}{\begin{eqnarray}} \nc{\eeqna}{\end{eqnarray}}

\nc{\lsq}{\left[} \nc{\rsq}{\right]} \nc{\lbc}{\left \{ }
\nc{\rbc}{\right \} } \nc{\lp}{\left(} \nc{\rp}{\right)}

\nc{\imp}{\Rightarrow} \nc{\lbf}{\lim_{b \rightarrow \infty}}
\nc{\limNinf}{\lim_{N \rightarrow \infty}} \nc{\limminf}{\lim_{m
\rightarrow \infty}} \nc{\limninf}{\lim_{n \rightarrow \infty}}
\nc{\convd}{\stackrel{\text{D}}{\longrightarrow}}
\nc{\convp}{\stackrel{\text{P}}{\longrightarrow}}
\nc{\convqm}{\stackrel{\text{qm}}{\longrightarrow}}
\nc{\eqd}{\stackrel{{\EuScript D}}{=}}
\nc{\convas}{\stackrel{a.s.}{\longrightarrow}}
\nc{\subi}{_{\text{I}}} \nc{\subs}{_{\text{S}}}
\nc{\subni}{_{\text{NI}}}
\nc{\trans}{^\top} \nc{\ol}{\overline}
\nc{\Ef}{ {\rm E}_{\infty} } \nc{\Ex}{ {\rm E} } \nc{\Ec}{ {\rm E}_1
} \nc{\Pf}{ {\rm P}_{\infty} } \nc{\Pc}{ {\rm P}_{1} } \nc{\Prb}{
{\rm P} } \nc{\sd}{\pm \hat{\sigma} }

\nc{\cond}{{\, \vert \,}} \nc{\indep}{{\, \perp \! \! \! \perp  \,}
} \nc{\tsps}{^{ {\rm T} } }

\nc{\pu}{\pi_{\rm U}} \nc{\pbi}{\pi_{\rm B}} \nc{\pnb}{\pi_{\rm NB}}
\nc{\prp}{\propto} \nc{\pr}{ {\rm pr} }
\nc{\half}{ {\textstyle \frac{1}{2}} }

\title[]{Causal comparative effectiveness analysis of dynamic continuous-time treatment initiation rules with sparsely measured outcomes and death}

\author{Liangyuan Hu$^{1*}$\email{liangyuan.hu@mssm.edu} and  
Joseph W. \ Hogan$^{2}$ \\
$^{1}$Icahn School of Medicine at Mount Sinai, New York, New York 10029, USA\\
$^{2}$Brown University School of Public Health, Providence, Rhode Island 02912, USA
}

\begin{document}

\date{{\it Received June} 2018. {\it Revised December} 2018.  {\it
Accepted December} 2018.}



\pagerange{\pageref{firstpage}--\pageref{lastpage}} 
\volume{64}
\pubyear{2018}
\artmonth{December}


\doi{10.1111/j.1541-0420.2005.00454.x}


\label{firstpage}


\begin{abstract}
Evidence supporting the current World Health Organization recommendations of early antiretroviral therapy (ART) initiation for adolescents is inconclusive. We leverage a large observational data and compare, in terms of mortality and CD4 cell count, the dynamic treatment initiation rules for HIV-infected adolescents. Our approaches extend the marginal structural model for estimating outcome distributions under dynamic treatment regimes (DTR), developed in Robins et al. (2008), to allow the causal comparisons of both specific regimes and regimes along a continuum. 
Furthermore, we propose strategies to address three challenges posed by the complex data set:
continuous-time measurement of the treatment initiation process; sparse measurement of
longitudinal outcomes of interest, leading to incomplete data; and censoring due to dropout
and death.  We derive a weighting strategy for continuous time treatment
initiation; use imputation to deal with missingness caused by sparse measurements and dropout; and define a composite outcome that incorporates both death and CD4 count as a basis for comparing treatment regimes.   Our analysis suggests that immediate ART initiation leads to lower mortality and higher median values of the composite outcome, relative to other initiation rules.  
\end{abstract}


%



\maketitle


%
\section{Introduction}
\subsection{Dynamic treatment regimes and treatment of pediatric HIV infection}

HIV/AIDS continues to be one of the leading
causes of burdensome disease in adolescents (10--19 years old).
Globally, an estimated 2.1 million adolescents were living with HIV in 2013, 
with most living in sub-Saharan Africa \citep{WHO2015}. Current World Health Organization (WHO) treatment recommendations for adolescents call for initiation of antiretroviral therapy (ART) upon diagnosis
with HIV \citep{WHO2015}.  Previously, and particularly for
resource-limited settings (RLS), WHO recommendations called for
delaying treatment until a clinical benchmark signaling disease
progression was reached.  For example, the 2013 guidelines recommended initiating ART when CD4 cell count -- a marker of immune system function -- fell below 500.

For investigating the effectiveness of ART initiation rules,  adolescents are a subpopulation of particular interest, particularly because of issues related to drug adherence \citep{mark2017hiv}. For adolescents, early initiation of ART can potentially increase the risk of poor adherence, leading to development of drug resistance, while initiating too late increases mortality and morbidity associated with HIV. 
Evidence from both clinical trials \citep{luzuriaga2004trial, violari2008early} and
observational studies \citep{berk2005temporal, schomaker2017optimal} supports the immediate ART initiation rule recommended by the WHO  for children under 10 years of age. 
Conclusive evidence is lacking for adolescents.  The 2015 WHO guidelines did not identify any study investigating the clinical outcomes of adolescent-specific treatment initiation strategies \citep{WHO2015}.   A recent large-scale study \citep{schomaker2017optimal} of HIV-infected children  (1--9 years) and adolescents (10--16 years) found mortality benefit associated with immediate ART initiation among children, but inconclusive results for the adolescents, 
and recommended further study of this group. Evaluating ART initiation rules specific to adolescents therefore remains important. 

Prior to 2015, WHO guidelines for treatment initiation were
expressed in the form of a dynamic treatment regime (DTR),
formulated as  ``initiate when a specific marker
crosses threshold value $q$''.  In a DTR,  the decision to initiate treatment for an
individual can depend on evolving treatment, covariate, and marker history 
\citep{chakraborty2014}. 
 
In this paper, we use observational data on 1962 HIV-infected adolescents, collected
as part of the East Africa IeDEA Consortium \citep{egger2012cohort}  to compare the effectiveness of
CD4-based DTR, with emphasis on comparisons  to
the strategy of immediate treatment initiation.  Our approach is to emulate a clinical
trial in which individuals are randomized at baseline and then followed for 
for a fixed amount of time, at which point mortality status and, for those remaining alive,
CD4 cell count are ascertained.  Hence the utility function for our comparison involves
both mortality and CD4 count among survivors.  

In addition to the usual complication of time-varying confounding caused by
treatment not being randomly allocated, the structure of the dataset poses
three specific challenges that we address here.  First, unlike with many
published analyses comparing dynamic treatment regimes, treatment initiation is
measured in continuous time; second, the outcome of interest, CD4, is measured
infrequently and at irregularly spaced time intervals, leading to incomplete
data at the target measurement time; third, some individuals may not complete
follow up, leading to censoring of both death time and CD4 count.

We use inverse probability weighting (IPW) to handle
confounding, and imputation to address missingness due to sparse measurement
and censoring.  To deal with continuous-time measurement
of treatment initiation, we derive continuous-time versions of the relevant
probability weights.  To deal with missingness, we rely on imputations from
a model of the joint distribution of CD4 count and mortality fitted to the observed
data.  We take a two-step approach:  first, the joint model is fitted to the
observed data and used to generate (multiple) imputations of missing CD4 and
mortality outcomes; 
second, we apply IPW to the filled-in datasets to generate causal comparisons
between different DTR.

\subsection{Comparing dynamic treatment regimes using observational data}
Randomized controlled trials can be used to evaluate a DTR of the form described above. 
(see \citealp{violari2008early} for example). Observational data afford large sample sizes and rich information on treatment decisions, but the lack of randomization motivates the need to use specialized methods for drawing valid causal comparisons
between regimes.
Statistical methods for drawing causal inferences about DTR from observational data include
the g-computation algorithm \citep{robins1986}, inverse
probability weighted estimation of marginal structural models \citep{robins2008estimation}, and
g-estimation of structural nested models \citep{moodie2007demystifying};
see \citet{daniel2013methods} for a comprehensive review and comparison.  

The g-computation formula was first introduced by
\citet{robins1986} and has been used to deal with time-dependent
confounding when estimating the causal effect of a time-varying
treatment.  The unobserved potential outcomes and
intermediate outcomes that would have been observed under
different hypothetical treatments are predicted from models for
potential outcomes and models for time-varying confounders. The predicted potential outcomes under different hypothetical DTR assignments are then contrasted for causal effect estimates.  
As the number of longitudinal time points increase, the method more heavily leverages parametric
modeling assumptions used for extrapolation of covariates and outcomes, increasing the
reliance on these assumptions and introducing potential for bias from model mis-specification.

The IPW approach re-weights each individual inversely by the probability of following specific regimes so that, 
in the weighted population, treatment can be regarded as randomly allocated to these regimes. 
Time-varying weights are required for handling time-dependent confounding. This involves specifying a model for treatment trajectory over longitudinal follow-up that can include time-dependent covariates. The IPW approach does not require models for 
the distribution of outcomes and covariates, which in principle makes it less susceptible to model misspecification than the g-computation formula.  The method can however generate  
unstable parameter estimates if there are extreme weights, raising the possibility of finite-sample
bias, which can often be alleviated by using stabilized weights or truncation 
\citep{cole2008constructing, cain2010start}.

\subsection{IeDEA data}
\begin{figure}[H]
  \centering
    \includegraphics[scale=0.6]{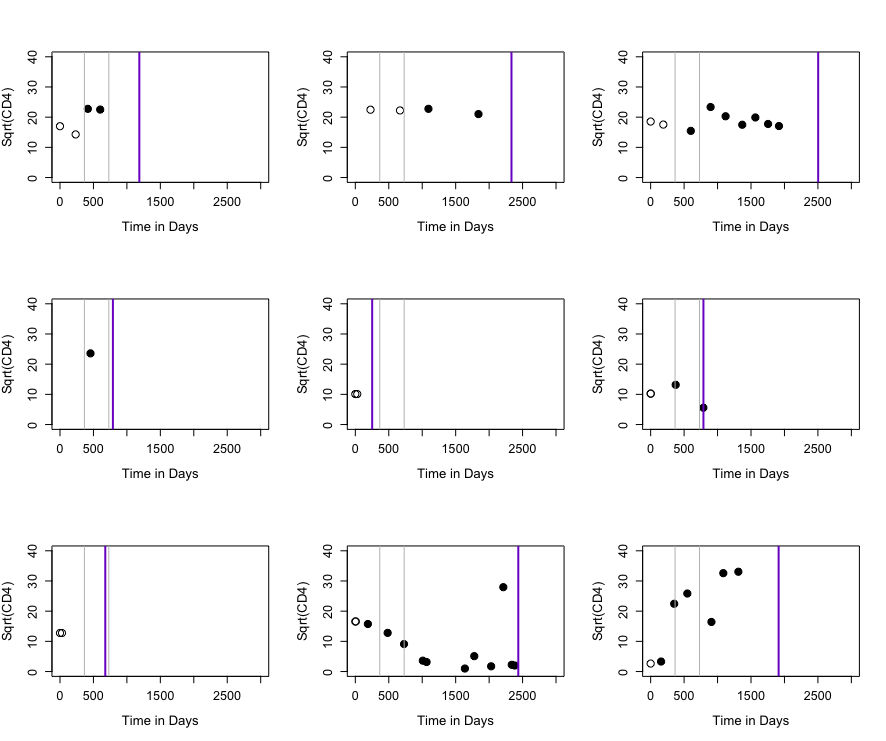} \caption{CD4 and ART initiation status during follow up 
for 9 randomly selected individuals. Empty circles indicate no ART and filled circles represent on ART.  Two gray lines denote one year and two years post diagnosis. Purple line corresponds to end of follow up.} \label{fig:CD4trajectory}
\end{figure}

The IeDEA consortium, established in 2005, collects clinical and demographic data
on HIV-infected individuals from seven global regions, four of which are in Africa. 
Data from African regions derive from 183 clinics providing ART \citep{egger2012cohort}. 
Our analysis makes use of clinical encounter data, drawn from the East Africa region,  on 1962 HIV-infected and ART naive adolescents who were diagnosed with HIV between February 20, 2002 and November 19, 2012. The dataset contains individual-level information at diagnosis on the following variables: age, gender, clinic site, CDC class (a 4-level ordinal diagnostic indicator of HIV severity), CD4 count, weight-for-age Z scores (WAZ) and height-for-age Z scores (HAZ). 
The dataset also includes longitudinal information on ART initiation status, death, CD4 count, WAZ and HAZ. These data were generated before the 2015 WHO guidelines that recommend immediate ART initiation,
which
yields significant variability in ART initiation patterns observed in our data. \
The follow-up visits vary considerably from patient to patient, resulting in irregularly and sparsely measured CD4 cell count (1.71, 1.32, 1.10  per person per year within one, two and three years of diagnosis) and various ART initiation patterns (Figure~\ref{fig:CD4trajectory}). 
Kaplan-Meier estimates of mortality  one-, two- and three-years post diagnosis are 3.3\%, 4.5\% and 5.6\% respectively. 

Our goal is to compare CD4 cell count and mortality rate at one and two years
post enrollment under dynamic regimes defined in terms of initiating treatment 
at specific CD4 threshold values.  In the next section we define the randomized 
trial our analysis is designed to emulate, and the outcome measure (utility) used for the
comparisons.

The remainder of the paper is organized as follows:  Section 2 describes notation and the statistical problem.  Section 3 delineates the approaches to estimating and comparing dynamic continuous-time treatment initiation rules with sparsely measured outcomes and death. Section 4 presents results from  our analysis of IeDEA data and highlights new insights relative to  previous studies.  Section 5 provides a summary and directions for future research.

\section{Notation and dynamic regimes}
\label{notation} 
\subsection{Randomized trial being emulated to compare dynamic regimes}
Ideally, causal comparisons of dynamic regimes should
be based on a hypothetical randomized trial \citep{hernan2006comparison}.
In our setting, the trial we are emulating would randomize
individuals at time $t=0$ to regimes in a set $\mathscr{Q} 
= \{0, 200, 210, 220, \ldots, 490, 500, \infty\}$, 
where $q=0$ corresponds to `never treat' and $q=\infty$ denotes
`treat immediately', and other regimes correspond to initiating
treatment when CD4 falls below $q$.   
Each individual would be followed to a specific time point $t^*$,
at which point survival status would be ascertained and, for those
surviving to $t^*$, CD4 would be measured.  For those who discontinue
follow up prior to $t^*$, we assume treatment status (on or off)
at the time of discontinuation would still apply at $t^*$.  

For each individual, let $\{ D_q : q \in \mathscr{Q} \}$ represent
the set of potential outcomes, one for each regime, indicating death at $t^*$,
such that $D_q=1$ if dead and $D_q=0$ if alive.  Similarly define
$\{ Y_q : q\in \mathscr{Q} \}$ to be the set of potential CD4 counts
for an individual who survives to $t^*$.  Now define, for $q \in \mathscr{Q}$,
the composite outcome $X_q = (1-D_q)Y_q$, with $X_q=0$
for those who die prior to $t^*$ and $X_q=Y_q > 0$ for those who survive.
We use both mortality rate $P(D_q=1) = P(X_q=0)$ and
quantiles of $X_q$ 
as a basis for comparing treatments.  The cumulative distribution
function (CDF) of $X_q$ is a useful measure of 
treatment utility because it has point mass at zero corresponding 
to the mortality rate, and thereby reflects information
about both mortality and CD4 cell count among survivors;
e.g., $P(X_q > 0)$ is the survival fraction and 
$P(X_q > x)$, for $x>0$, is proportion of individuals who survive
to $t^*$ {\it and} have CD4 count greater than $x$.

\subsection{Defining dynamic treatment regime}
Let $\{Z(t): t\geq 0\}$, where $Z(t)>0$,
represent CD4 cell count, which is defined for all $t$ but measured
only at discrete time points for each individual (see below).
Let $T$ denote survival
time, with $\{ N^T(t) : t>0 \}$ its associated zero-one counting 
process.  Each individual has a  $p \times 1$ covariate process
$\{ L(t) : t \geq 0  \}$, some elements of which may be time varying. 
The time-varying covariates may be recorded at times other than
those where $Z$ is recorded.
Finally let $A$ denote the time of treatment initiation,
with associated counting process $\{ N^A(t) : t \geq 0 \}$ and intensity function $\lambda^A(t)$. 
Adopting a convention in the DTR literature \citep{robins2008estimation}, we assume the decision 
to initiate ART at $t$ is made after observing the covariates and CD4 cell count; that is,
for a given $t$,  $N^A(t)$ occurs after $Z(t)$ and $L(t)$.   Finally let $C$ be a censoring (dropout) time, with
associated counting process~$N^C(t)$.

At a fixed time $t$, let $H(t) = \{Z(t), N^T(t), L(t), N^A(t), N^C(t)\} $ represent
the most recent values of each process.  We use overbar notation to denote
the history of a process, so that (e.g.) $\ol{L}(t) = \{ L(s): 0 \leq s \leq t \}$
is the history of $L(t)$ up to $t$. All individuals are observed at baseline and then at
a discrete number of time points whose number, frequency and spacing may vary.
Hence the observed-data process for individual $i$ $(=1,\ldots,n)$ is denoted by
$\ol{H}_i(t_{iK_i}) = \{ H_i(t) : t = 0, t_{i1}, t_{i2}, \ldots, t_{iK_i} \}$.

\nc{\Zmin}{Z_{\min}}
\subsection{Mapping observed treatment to dynamic treatment regime}
\label{sec:rules}
The dynamic treatment regime `initiate treatment when $Z(t_j)$ falls below threshold $q$' (where $t_j$ is time at the $j$th visit) is a deterministic function $r_q(\ol{H}(t_j))$ that depends on observed values of 
$\ol{Z}(t_j)$ and treatment history $\ol{N}^A(t_j)$; for brevity we suppress subscript $j$ and write $r_q(t)$, which applies to each individual's actual visit times.  As some patients have missing  baseline CD4, let $R^Z(t)$ be a binary indicator with $R^Z(t)=1$ denoting that CD4 has not been observed by time $t$. At $t=0$, the rule is $r_{q}(0) = I \{R^Z(0) =1  \text{ or } Z(0) < q  \} $, indicating immediate initiation regardless of $Z(0)$ or treat if $Z(0)$ is below $q$. 
For $t>0$, we define $Z_\text{min} (t) = \min_{j: 0 \leq t_j < t} Z(t_j)$ to be the lowest previously recorded value of $Z$ prior to $t$.
 Then, 
	\beq
	r_{q}(t) & = & 
	\lbc \begin{array}{cl}
			 0 & \text{ if }   \{ N^A(t^-) = 0 \text{ and }  Z_{\min}(t)  \geq q \text{ and } Z(t) \geq q \} \text{ or } R^Z(t) =1,\\
	           1 & \text{ if } N^A(t^-) = 0 \text{ and } Z_{\min}(t) \geq q \text{ and } Z(t) < q, \\
	           1 & \text{ if } N^A(t^-) = 1.
	\end{array}
	\right.
\eeq

In words, the first line of the rule says not to treat if an individual has not yet 
initiated treatment and $Z(t)$ has not fallen below $q$ or has not been observed; the second line says to
treat if time $t$ represents the first time $Z(t)$ has fallen below $q$; the third
line says to keep treating once ART has been initiated.

In addition to the observed data process, we define a regime-specific 
compliance process $\{ \Delta_{q}(t): t\geq 0 \}$ where $\Delta_q(t)=1$
if regime $q$ is being followed at time $t$ and $\Delta_q=0$ otherwise.
Written in terms of $\ol{H}(t)$ and $r_q(t)$, we have
$\Delta_q(t) = N^A(t) r_q(t) + \{ 1- N^A(t) \} \{ 1 - r_q(t) \}.$
Hence if an individual's actual treatment status at time $t$ agrees with the DTR $q$, then this individual is compliant with regime $q$ at time $t$.  Thus for each individual and for each $q \in \mathscr{Q}$, we observe, in addition
to $\ol{H}(t)$, a regime compliance process 
$\lbc \Delta_{qi}(t) : t=0, t_{i1}, \ldots, t_{iK_i} \rbc$.

\subsection{Missing outcomes due to sparse measurement times and censoring}
For those who remain alive at $t^*$, the observed  $X_i$
corresponds to  $Z_i(t^*)$. 
When measurement of $Z_i(t)$ is sparse and irregular,
 $Z_i(t^*)$ will not be directly observed unless $t_{ik} = t^*$ for some $k
 \in \{ 1, \ldots, K_i \}$.
In settings like this, it is common to 
define the observed outcome as the value of $Z_i(t)$ closest 
to $t^*$ and falling within a pre-specified interval $[t_a, t_b]$ 
containing $t^*$.  Specifically, $X_i$ is
the value of $Z(t_{ik})$ such that $t_{ik} \in [t_a, t_b]$ and  
$|t_{ik} - t^*|$ is minimized over $k$.  Even using this definition,
the interval $[t_a, t_b]$ still may
not contain any of the measurement times for some individuals; hence 
$X_i$ can be missing even for those
who remain in follow up at $t^*$.
The other cause of missingness in $X_i$ is dropout,
which occurs when $t_{iK_i} < t_a$.  

For both of these situations,
we  rely on multiple imputation based on a model for the 
joint distribution of the CD4 process $Z(t)$ and the mortality
process $N^T(t)$.  The general strategy is as follows:
first, we specify and fit a model for the joint distribution
$[Z(t), N^T(t) \cond \ol{H}(t)]$ of CD4 and mortality, conditional
on observed history.
For those who are known to be alive but do not have a CD4 measurement
within the pre-specified interval $[t_a, t_b]$, we 
impute $\wt{X}_i \sim  [ {Z}(t^*) \cond \ol{H}_i(t^*) ] $ from
the fitted CD4 submodel.  For those who are missing $X_i$ 
because of right censoring, we proceed as follows:  (i)~calculate 
$P\{ N^T(t^*) = 1 \cond \ol{H}_i(t_{iK_i}) \}$ from the fitted survival submodel, and impute
$\wt{D}_i$ from a Bernoulli distribution having this probability;
(ii)~for those with $\wt{D}_i=0$, impute $\wt{X}_i \sim  [ {Z}(t^*) \cond \ol{H}_i(t^*) ] $ from
the fitted CD4 submodel; (iii)~for those with $\wt{D}_i=1$, set $\wt{X}_i=0$.  
Further details are given in Section~\ref{sec:imp}.

\section{Estimating and comparing effectiveness of dynamic regimes}
\label{sec:com_eff}

\subsection{Assumptions needed for inference about dynamic regimes}
We are interested in parameters or functionals of
the potential outcomes distribution $F_{X_q}(x) = P( X_q \leq x)$.
Specific quantities of interest are the mortality rate 
$\theta_{q1} = P(X_q = 0) = F_{X_q}(0)$, the median of the distribution of the composite outcome $\theta_{q2} = F_{X_q}^{-1}(\half)$,
and the mean CD4 count among survivors $\theta_{q3} = E(X_q \cond X_q > 0)$. 
We first consider
inference in the
case where there is no missingness in the observable outcomes $X_i$. 
Estimates for each of these quantities can be obtained using
weighted estimating equations under specific assumptions.

\nc{\Hbar}{\ol{H}}
{\emph {A1. Consistency assumption}.}
To connect observed
data to potential outcomes, we use the {\emph consistency relation} 
$X_i =  X_{qi}$ when $\Delta_{qi}(t^*)=1$, for all $q \in \mathscr{Q}$, which implies that the observed outcome $X_i$ 
corresponds to the potential outcome $X_{qi}$ when individual $i$ actually follows
regime $q$.  Note that an individual can potentially follow more than one
regime at any given time.

{\emph {A2. Exchangeability assumption}.}
In observational studies, individuals are not randomly assigned to
follow regimes. Decisions on when to start ART are often made by
based on guidelines and observable patient characteristics.  We make the following exchangeability
assumption, also known as sequential randomization of treatment:
$\lambda^A(t \cond \Hbar(t), T>t, X_q) =  \lambda^A(t \cond \Hbar(t), T>t)$,
for $t<t^*$. 
This assumption states that initiation of treatment at $t$ among those
who are still alive is conditionally independent of the potential outcomes
$X_q$ conditional on observed history~$\Hbar(t)$.

{\emph {A3. Positivity assumption}.}
Finally we assume that at any given time $t$, there is positive probability
of initiating treatment, among those who have not yet initiated, for all configurations $\Hbar(t)$
\citep{robins2008estimation}:
$P \lbc \lambda^A(t \cond \ol{H}(t), T>t) > 0 \rbc = 1$.  This implicitly assumes a positive probability of visiting clinic in the interval $[t,t^*]$, conditional on $\ol{H}(t)$.

\subsection{Weighted estimating equations for comparing specific regimes}
For illustration, consider estimating
the mortality rate $\theta_{q1} = P(X_q=0)$.  If
individuals are randomized to specific regimes, a consistent estimator
of the death rate is the sample proportion among those who follow regime $q$; i.e., $\wh{\theta}_{q1} = \sum_i \Delta_{qi}(t^*) 
I ( X_i = 0 )  
\left/ \sum_i \Delta_{qi}(t^*) \right.$.  This estimator is 
the solution to $\sum_i \Delta_{qi}(t^*) \{ I ( X_i = 0 ) - \theta_{q1}\} = 0$, which is 
an unbiased estimating equation when $\theta_{q1}=\theta_{q1}^*$ is
the true value of $\theta_{q1}$.  We can similarly construct unbiased estimating
equations for other quantities of interest.  For example, under randomization, a
consistent estimator of the median
of $X_q$ is the solution to $\sum_i \Delta_{qi}(t^*) \lbc I(X_i \leq \theta_{q2}) - \half \rbc = 0$.

For observational data, 
relying on the assumptions of consistency, positivity and exchangeability,
we can obtain consistent estimates of quantities of interest
using weighted estimating equations.
Returning to mortality rate,
a consistent estimator of $\theta_{q1}$ can be obtained as the solution to the
weighted estimating equation
$\sum_{i=1}^n \Delta_{qi}(t^*) W_{qi} \lbc I(X_i = 0) - \theta_{q1} \rbc =0$,
where $W_{qi} = 1 / P \{ \Delta_{q}(t^*)=1 \cond \Hbar_i(t^*) \}$
is the inverse probability of following regime $q$ through
time~$t^*$ \citep{robins2008estimation, cain2010start, shen2017estimation}.

In practice the weights $W_{qi}$ must be estimated from data;
some of the estimated weights can be large, leading
to estimators with high variability \citep{cain2010start}. This problem can be ameliorated to some degree by using stabilized weights of the form
\beqna
\label{eq:sipw}
W^s_{qi} &=& \frac{P \lbc \Delta_{q}(t^*)=1 \rbc}{P \lbc \Delta_{q}(t^*)=1 \cond \Hbar_i(t^*) \rbc }.
\eeqna
In this case, the numerator of the weight function needs to be calculated
directly from the regime indicator processes.  Specifically, for each
regime $q$, define a 0-1 counting
process $N^q(t) = 1-\Delta_q(t)$ that jumps when regime $q$ 
is no longer being followed, and let $\Lambda^q(t)$ denote its associated
cumulative hazard function.  Then $S^q(t) = P \{ N^q(t)=0 \} = P \{ \Delta_q(t) = 1 \}$;
hence (an estimate of) $S^q(t^*) = \exp\{ -\Lambda^q(t^*) \}$ can be used as the numerator weight.

\subsection{Comparing regimes along a continuum}\label{sec:MSM}
We can examine the effect of DTR $q$ on $X_q$ at a higher resolution along a continuum such as $\mathscr{Q} = \{200,210,\ldots,500 \}$ (we use integers for $\mathscr{Q}$, but theoretically it can include continuous values). When the number of regimes to be compared is large, it is highly possible that not every regime is followed by a sufficiently large number of
individuals, and sampling variability associated with the regime effect estimated using the procedure for discrete regimes may be large  \citep{hernan2006comparison}.
A statistically more efficient approach is to formulate a causal
model that captures the smoothed effect of $q$ on a parameter of interest; we illustrate
using the median  $\theta_{q2} = F^{-1}_{X_q}(\half)$. 

Let $q_l$ and $q_u$ denote the lower and upper bound of the regime continuum. Assume $F^{-1}_{X_q}(\tau)$, where $\tau$ is a fixed quantile, follows a structural model
\beqna
\label{eq:mod_b} F^{-1}_{X_q}(\tau)  &=& \alpha_0 I(q = \infty) + \alpha_1 I(q = 0) + I( q \in [q_l, q_u] ) d(q) ,
\eeqna
where $d(\cdot)$ is an unspecified function with smoothness constraints. In our application, 
we use natural cubic splines constructed from piecewise third-order
polynomials that pass through a set of control points, or knots,
placed at quantiles of $q$. This  allows $d(q)$ to flexibly capture the effect of $q$ along the continuum and enables separate estimation of the discrete regimes $q = \infty$ and~$q=0$. 
Parameterizing our model in terms of the basis functions of a natural cubic spline with $J$ knots \citep{hastie2009elements} yields 
$F^{-1}_{X_q}(\tau) = \alpha\trans V(q)$, where
$$V(q)_{(J+2)\times 1} = [I(q=\infty), \; I(q=0), \;I( q \in [q_l, q_u] ) d^\dagger(q) \trans] \trans $$ and $d^\dag(q) = [ d^\dag_1(q), \cdots, d^\dag_J (q)
]\trans$ are the $J$ basis functions of $d(q)$. The parameter $\alpha$ is a vector of $J+2$ coefficients for $I(q = \infty)$, $I(q=0)$ and the basis
functions $d^\dag(q)$. The causal effect of regime $q$ on the potential outcome $X_q$ is therefore
encoded in the parameter $\alpha$. A consistent estimator of $\alpha$ can be obtained by solving the estimating equation \citep{Leng2014}:
\beq
\sum_i \Delta_{qi}(t)W^s_{qi} V_i(q) \lsq I\{X_i - V_i\trans(q)\alpha > 0\} - \tau \rsq&=& 0.
\eeq
Setting $\tau = 0.5$ estimates the causal effect of $q$ on the median of  $X_q$.  


\subsection{Derivation and estimation of continuous time weights}
\label{sec:dev_ipw}
\subsubsection{Assuming no dropout or death prior to $t^*$}
\label{subsec:nodropout}
The denominator of $W^s_{qi}$ in equation~\eqref{eq:sipw} is the probability of individual $i$ following regime $q$ through $t^*$, conditional on observed history $H_i(t^*)$.
As described in \cite{robins2008estimation, cain2010start} and \cite{shen2017estimation},  
for discrete-time settings where the measurement times are common across individuals, this probability corresponds to the cumulative product of conditional probabilities of treatment indicators over a set of time intervals $0 = t_{0} < t_2< \cdots <t_{K}=t^*$. 
Specifically,  
\beqna
P \lbc \Delta_{qi}(t^*)=1 \; \left| \; \ol{H}_i(t^*)\rbc \right. \nonumber 
&=& \prod_{k=0}^{K} P \lp \Delta_{qi}(t_{k})=1 \; \left| \; \ol{H}_i(t_{k})\rp \right. \nonumber\\
&=& \prod_{k=0}^{K} P \lsq \lbc N^A_i(t_{k})r_{qi}(t_{k}) + (1-N^A_i(t_{k}))(1-r_{qi}(t_{k}))\rbc  = 1 \, \left| \, \ol{H}_i(t_{k}) \rsq \right. \nn\\
&=& \prod_{k=0}^{K}  \lbc P\left(N^A_i(t_{k})=1 \, \left| \, \ol{H}_i(t_{k})\right)I\lp r_{qi}(t_{k}) = 1\rp \right. \right. \nn \\
& & \hspace{.5in} + \left. P\left(N^A_i(t_{k})=0 \, \left| \, \ol{H}_i(t_{k})\right) I \lp r_{qi}(t_{k}) = 0\rp \rbc \right. . \label{eq:prodint}
 \eeqna
This establishes the connection between regime compliance and treatment history. Equation~\eqref{eq:prodint} represents the treatment history among those with $\Delta_{qi}(t^*) =1$; therefore, to compute the probability of regime compliance for those with $\Delta_{qi}(t^*) =1$, we just need to model their observed treatment initiation process, as described in equation~\eqref{eq:prodint1} below.   

This observation allows us to  generalize the weights for the discrete time setting to the continuous time process. Let $dN_i^A(t)$ be the increment of $N_i^A$ over the small time interval $[t, t+dt)$.  Note that conditional on $\ol{H}(t)$, the occurrence of treatment initiation for individual $i$ in $[t, t+dt)$ is a Bernoulli trial with outcomes $dN_i^A(t) = 1$ and $dN_i^A(t) =0$. Equation~\eqref{eq:prodint} can therefore be written 
\beqna
\label{eq:prodint1}
\prod_{k=0}^K P \lp dN_i^A(t) =1 \cond \ol{H}_i(t) \rp^{dN_i(t)}  P \lp dN_i^A(t) =0 \cond \ol{H}_i(t) \rp^{1-dN_i(t)},
\eeqna
which takes the form of the individual partial likelihood for the counting process $\{N_i^A(t): 0\leq t \leq t^*\}$. When the number of time intervals between $t_0$ and $t_K$ increases, $dt$ becomes smaller, 
and the finite product in~\eqref{eq:prodint1} will approach a product-integral \citep{aalen2008survival}
\beqna
\Prodi_{0 \leq t \leq t^*} \lbc \lambda^A(t \cond  \Hbar_i(t) )dt \rbc^{dN^A_i(t)} 
\lbc 1- \lambda^A(t\cond  \Hbar_i(t) )dt \rbc^{1-dN^A_i(t)}\label{eq:prodint2}
 \hspace{2in} \\
 = \lsq \prod_{0 \leq t \leq t^*} \lbc \lambda^A(t\cond  \Hbar_i(t)) \rbc^{\Delta N^A_i(t)} \rsq \exp \lbc -\int_0^{t^*}  \lambda^A(t\cond  \Hbar_i(t))dt   \rbc, \hspace{1in}
 \label{eq:prodint3}
\eeqna
where $\Delta N^A_i(t) = N^A_i(t) - N^A_i(t^-)$. The product integral of the first part in~\eqref{eq:prodint2} is the finite product over the jump times of the counting process, hence the first factor in~\eqref{eq:prodint3}. The second factor in~\eqref{eq:prodint3} follows from properties of the product-integral of an absolutely continuous function (\citealt{aalen2008survival}, Appendix A.1). 

The individual counting process $\{N_i^A(t), 0 \leq t \leq t^*\}$ will have at most one jump (at $A_i$), and in our case patients stay on ART once it is initiated.
Hence the product integral only needs to be evaluated up to the ART initiation time.
Equation~\eqref{eq:prodint3} therefore reduces to
\beqna
P\lbc\Delta_{qi}(t^*) = 1 \cond \Hbar_i(t^*)\rbc  &=&  \lambda^A(A_i \cond \Hbar(A_i)) S^A(A_i \cond \Hbar_i(A_i)) N_i(t^*) 
+ S^A(t^* \cond \Hbar_i(t^{*})) \{ 1-N_i(t^*) \}  \nonumber\\
		&=& f^A(A_i \cond \Hbar_i(A_i)) N^A_i(t^*) + S^A(t^* \cond \Hbar_i(t^{*})) \{ 1-N^A_i(t^*) \},\label{eq:den_wt_cont}
\eeqna
where $S^A(t \cond \Hbar(t)) = \exp\{ -\Lambda^A(t \cond \Hbar(t)) \}$ is the 
survivor function associated with the ART initiation process.  


For an alternate derivation of the continuous time weights, see \cite{johnson2005semiparametric}, who use a  Radon-Nikodym derivative of one integrated intensity process (under randomized treatment allocation) with respect to another (for the observational study), and arrive at the same weighting scheme as ours.  Simulation studies by \cite{hu2017modeling} demonstrate consistency and
stability of weighted estimators using continuous-time weights in empirical settings
when assumptions A1-A3 hold and the weight model is correctly specified.  

Components of the denominator weights are 
estimated from a fitted hazard model for treatment
initiation.  Specifically we assume   $\lambda^A(t \cond \Hbar(t))$ follows
a Cox proportional hazards model $\lambda^A(t \cond \Hbar(t)) = 
\lambda_0^A(t) u(\Hbar(t) ; \phi)$, where $u$ is a strictly positive 
function capturing the effect of covariates and $\phi$ is a finite-dimensional
parameter vector.  Details of the model specification
used in our application are given in Section~\ref{sec:application}.  The parameter $\phi$  is estimated
using maximum partial likelihood estimation, and the baseline hazard function $\lambda^A_0(t)$
is estimated using the Nelson-Aalen estimator.  The functions $f^A$ and $S^A$ are estimated via
\beqna
\wh{S}^A(t \cond \Hbar(t)) 
                      &=& \exp \lbc -\int_0^t  u( \Hbar(s); \wh{\phi}) \; d\wh{\Lambda}^A_0(s) \rbc,\label{eq:est_S}\\
\wh{f}^A(t \cond \Hbar(t)) &=& d\wh{\Lambda}^A(t \cond \ol{H}(t)) \; \wh{S}^A(t \cond \ol{H}(t)).\label{eq:est_f}
\eeqna
To estimate the stabilizing numerator weight $P\lbc \Delta_{q} (t^*) =1
\rbc$, we use the $q$-specific survivor function associated with the
counting process $N^q(t)$, estimated using the Nelson-Aalen estimator   
$\wh{S}^q(t^*) = \exp\{-\wh{\Lambda}^q(t^*)\}$.

\subsubsection{Considering dropout or death prior to $t^*$}
In the IeDEA data, some participants drop out prior to $t^*$,
which requires modifications to the weight specification.   We make an additional assumption: 

{\emph {A4. Conditional constancy assumption}.} Once lost to follow up at $C_i<t^*$, 
treatment and regime  status remain constant;  
i.e., $N^A(t) = N^A(C_i)$ and $\Delta_{qi}(t) = \Delta_{qi}(C_i)$
for all $t \in [C_i, t^*]$.

Under this assumption, both regime adherence and treatment initiation status
are deterministic after $C_i$.  Hence the stabilized weight is $S^q(C_i)/ f^A(A_i \cond \ol{H}(A_i))$
for those who initiated treatment prior to $C_i$ and
 $S^q(C_i)/ S^A(C_i \cond \ol{H}(C_i))$ for those who have not. 
If death occurs at $T_i<t^*$, both compliance and treatment initiation processes only need to be evaluated up to time $T_i$, and  estimation of the stabilized weights is same as described above,
with $T_i$ replacing~$C_i$.   
Let $U_i=\min(T_i,C_i,t^*)$ denote duration of follow up time
for individual $i$.  The modified stabilized weight can be written as
\beqna
W^s_{qi} &=& \dfrac{S^q(U_i)I(U_i <t^*) + S^q(t^*) I(U_i \geq t^*)}{f^A(A_i \cond \ol{H}(A_i))}N_i^A(t^*) \nonumber\\
&&+ \lbc \dfrac{S^q(U_i)}{S^A(U_i \cond \ol{H}_i (U_i))} I(U_i<t^*) +  \dfrac{S^q(t^*)}{S^A(t^* \cond \ol{H}_i (t^*))} I(U_i\geq t^*)\rbc \lp 1-N_i^A(t^*)\rp. \label{eq:swt_final}
\eeqna
Estimation follows by equations~\eqref{eq:est_S} and~\eqref{eq:est_f}.

\subsection{Imputation strategy for missing and censored outcomes}
\label{sec:imp}
Imputation of missing CD4 counts and mortality status are generated
from a joint model of CD4 and survival.  The two processes are linked
via subject-specific random effects that characterize the true CD4 
trajectory \citep{rizopoulos2012joint}.
Hazard of mortality is assumed to depend on the
true, underlying CD4 count as described below.

Observed CD4 counts as a function of time are specified with
a two-level model.  At the first level, $Z_i(t)   =  m_i(t) + e_i(t)$, 
where $m_i(t)$ is the true, underlying CD4 cell count and $e_i \sim N(0,\sigma(t))$ is
within-subject variation of the observed counts around the truth.  
The second level specifies the trajectory in terms of baseline covariates
$L_i(0)$, treatment initiation time $A_i$, follow up time $t$, and subject-specific
random effects $b_i$,
\beq
m_i(t)  &=&  h_1(L_i(0), N^A_i(t), t;  \beta) + h_2(N^A_i(t), t;  b_i).
\eeq
In the model for $m_i(t)$, 
$h_1(L_i(0), A_i, t;  \beta)$ models the effect of $L(0)$, $A$ and $t$
in terms of a population-level parameter $\beta$ and $h_2(A_i, t;  b_i)$
captures individual-specific time trajectories relative to treatment initiation in terms 
of random effects $b_i$, where where $b_i 
\sim  N(0, \Omega)$.

The hazard model for death uses true CD4 count $m_i(t)$ as a covariate,
in addition to components of $L_i(0)$ and treatment timing.  The specification
we use in our analysis is 
\beqna
\log \lambda^T(t \cond m_i(t), {L}_i(0), A_i(t) )
&=& \log \lambda^T_0(t) + g_1(m_i(t); \gamma_1) + g_2({L}_i(0), N^A_i(t); \gamma_2) 
\eeqna
where $\lambda^T_0(t)$ is an unspecified baseline hazard function,
$g_1(\cdot; \gamma_1)$ is a smooth, twice-differentiable function
indexed by a finite-dimensional parameter $\gamma_1$; 
and $g_2(\cdot; \gamma_2)$ captures the main effect of baseline covariates,
the instantaneous effect of treatment initiation, and potential interactions between them.
In our application, we use cubic smoothing splines
to model the effects of $m_i(t)$ and of continuous baseline covariates.
This model has fewer covariates than the CD4 model because of relatively
low mortality rates.    

The joint model is used to generate imputations where CD4 count and
mortality information are missing at time $t^*$.
The variance of our target parameters $\theta_q 
= (\theta_{q1}, \theta_{q2}, \theta_{q3})$ is based on 
Rubin's variance estimator \citep{rubin1987multiple};
full details of model specifications and variance calculations used
in the data analysis in Section 4 appear in 
Supporting Information.

%
%

\section{Application to IeDEA data}\label{sec:application}

Our analysis uses longitudinal data on 1962 adolescents with at least two years
of follow up time.
Time is measured in days.
We evaluate effectiveness of the regimes at times  $t^* = 365$ and  $t^* = 730$ days (one and two years, respectively) after diagnosis. To capture the CD4 observed at $t^*$,  we  set $[t_a, t_b] = [t^*-180, t^*+180]$; hence $Y$ is the CD4 count measured at a time falling within $[t_a, t_b]$ and closest to $t^*$.  If no CD4 is captured within $[t_a, t_b]$, then $Y$ is  missing. The percentage of missing data for $Y$ is  29.1\% at one year and 43.4\% at two years. Among those with missing one-year outcome, 41.2\% were lost to follow up prior to $t_a$;
for those with missing two-year outcome the proportion is 42.5\%.  Table~\ref{tab:baseline} describes summary statistics for baseline variables and follow up, the observed outcome pair
$(Y, D)$ (CD4 and deaths),  and ART initiation. 

Missing outcomes are imputed following the strategies described in Section~\ref{sec:imp}, and the complete datasets are analyzed using 
IPW methods for the causal comparative analysis.  The fit of the CD4 submodel was
examined using residual plots and examination of individual-specific fitted curves;
for the mortality submodel we tested the proportional hazards assumption for each
term included in the model.  These model checks indicated no evidence of lack of fit.
Details appear in Supporting Information.

Following the deterministic rule $r_q(\ol{H}( t ))$ described in Section~\ref{sec:com_eff}, we create the regime-specific indicators $\Delta_{qi} (t^*)$ for  $q \in \mathscr{Q}$  for each patient  based on the concordance between their ART initiation history $\{N^A(t): 0 \leq t \leq t^*\}$ and $r_q(\ol{H}(t^*))$.
To estimate regime weights, we fit the model $\lambda^A(t \cond \Hbar(t)) = 
\lambda_0^A(t) u(\Hbar(t) ; \phi)$ to  individuals' treatment and covariate histories observed in the original data to estimate the denominator of $W_{qi}^s$ in~\eqref{eq:sipw}.  For the time-varying
component of $\Hbar(t)$, we include the most recently observed values of CD4, WAZ and HAZ as main effects, modeled 
using cubic splines.  For baseline covariates, we include age at diagnosis (modeled using a cubic spline) and the categorical variables gender and CDC symptom classification (mild, moderate, severe, asymptomatic, missing). 
To estimate the numerator of the stabilized weights, we use the Nelson-Aalen estimator 
of the survival function for each regime-specific compliance process, as described 
in Section~\ref{subsec:nodropout}. We truncated the weights at 5\% and 95\% quantiles to improve stability. We conducted a sensitivity analysis to assess the impact of weight truncation. The point estimates and the confidence intervals for treatment effect on mortality were unchanged with different weighting schemes.  Point estimates and variation associated with treatment effect on the composite outcome increased with less truncation;  the
confidence intervals indicated greater variability but no change in substantive conclusion about treatment effect. For the denominator weight model, we tested the proportional hazards assumption for each term included in the model and found no violations of the assumption. Details appear in Supporting Information.

We summarize the comparative effectiveness for specific regimes $q \in \{0, 200, 350, 500,  \infty\}$ in Table~\ref{tab:res_dis}  in terms of mortality proportion $\theta_{q1} = P(X_q=0)$, median of the distribution of  the composite outcome $\theta_{q2} = F_{X_q}^{-1}(\half)$, and mean CD4 count among survivors, $\theta_{q3} = E(X_q \cond X_q>0)$. (The quantity  $\theta_{q3}$ is not a causal effect because it conditions on having survived to time~$t^*$.) Confidence intervals are constructed using the normal approximation to the sampling distribution, derived from bootstrap resampling, as described in Supporting Information.  
 
Immediate ART initiation yields significantly lower mortality rate and higher medians of the composite outcome at both years than delayed initiation.  The ``never treat'' regime  leads to significantly higher mortality rate; among the patients who survive to one year, CD4 is higher -- resulting in higher $\theta_{q2}$ and $\theta_{q3}$ -- indicating that those who do survive without treatment may be relatively healthier at the beginning of the follow up.
 
\begin{figure}[H]
  \centering
    \includegraphics[scale=0.75]{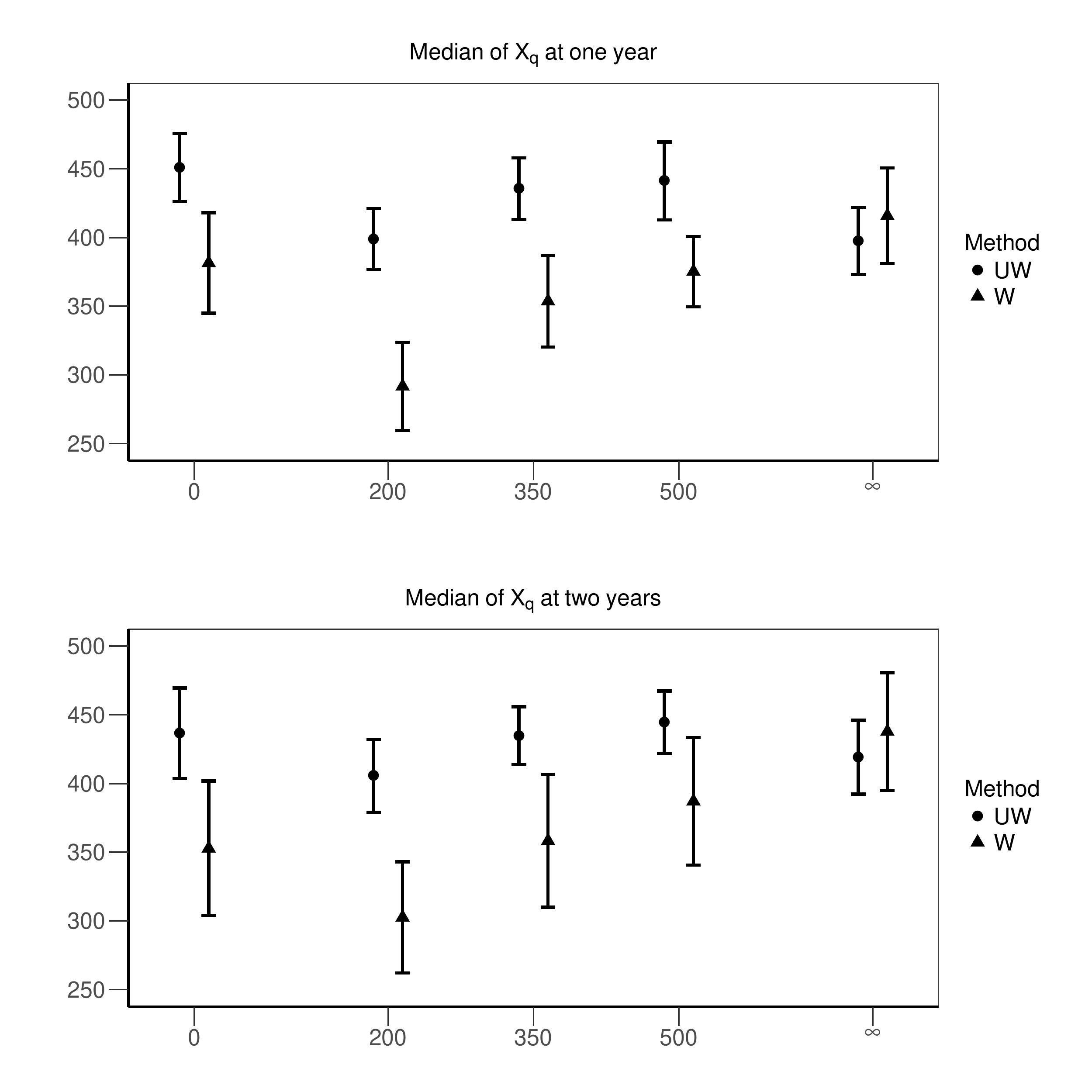} \caption{Comparing the median values of $X_q$ under regime $q \in \{0, 200, 350, 500, \infty\}$. Weighted (W) and unweighted (UW) estimates are compared side-by-side.}\label{fig:disc_regm}
\end{figure}

Figure~\ref{fig:disc_regm} shows the effect of weighting on  estimated medians of $X_q$
for  $q = 0, 200,350, 500,\infty$.    
We compare weighted and unweighted estimates using imputed data;  the weighted estimates suggest immediate ART initiation leads to highest $\wh{\theta}_{q2}$, whereas the unweighted estimates ignoring nonrandom allocation of DTRs recommend `never treat' to be the optimal regime. The difference could be attributable to differences in baseline covariates (see Table 8 in Supporting Information). Not surprisingly, the weighted estimates have
higher variability.  

\begin{figure}[H]
  \centering
  \includegraphics[scale = 0.72]{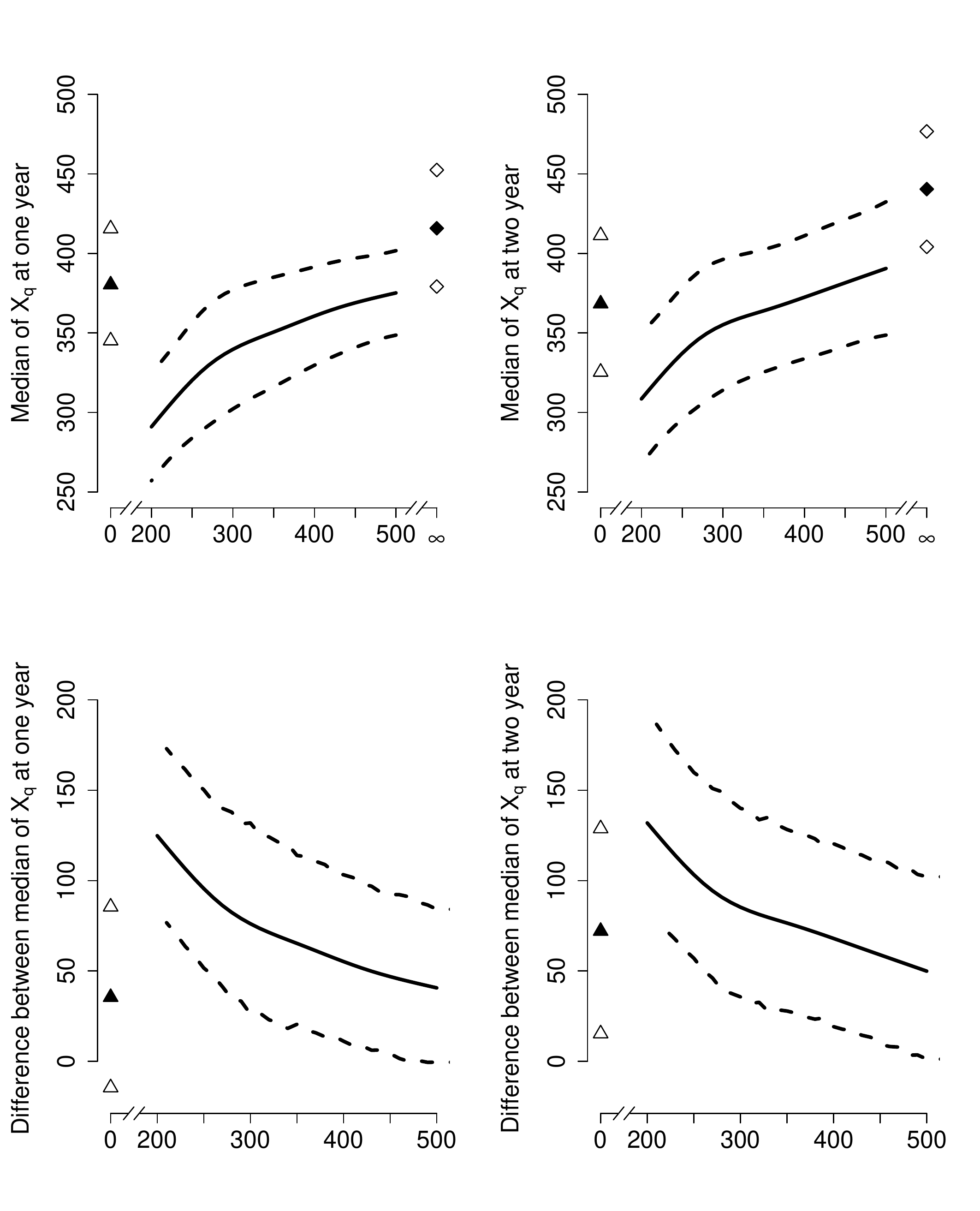}
  \caption{The effectiveness of continuous regimes. The upper panel presents the median of $X_q$, $\wh{\theta}_{q2}$, at one-year and two-years; the bottom panel displays the difference in $\wh{\theta}_{q2}$ at one-year and two-years between regimes $q = \infty$ and $q \in \{0, 200, 210,  \ldots, 500\}$. The triangles represent $\wh{\theta}_{q2}$ corresponding to regime $q=0$ (upper panel), and the difference in $\wh{\theta}_{q2}$ between regimes $q=\infty$ and $q=0$ (bottom panel). Similarly, the diamonds correspond to $\wh{\theta}_{q2}$ under regime $q=\infty$. The filled symbols are the mean values, and the empty symbols are the upper and lower bounds of the 95\% confidence intervals. } \label{fig:smooth_curv}
\end{figure}

Finally, we estimate the causal effect of the DTR on the median of $X_q$ 
using the smoothed relationship between $F^{-1}_{X_q}(\half)$ and $q$ from model~\eqref{eq:mod_b}. The estimated  ``dose response''
curves of $\wh{\theta}_{q2}$ versus $q$ appear in the top panel of
Figure~\ref{fig:smooth_curv}. The bottom panel describes the difference in $\wh{\theta}_{q2}$ between dynamic regimes $q=\infty$ and $q \in \{0, 200, 210, \ldots, 500\}$. Our results indicate that immediate ART initiation leads to significantly higher median values of the composite outcome $X_q$ than delayed ART initiation. Furthermore, as an illustration of increased efficiency, the variance of the one-year outcome associated with $q = 350$ estimated from the structural model  is 180, compared to 209 for the regime-specific estimate, a 13.9\% reduction. The \verb+R+ code used to implement our approaches is available in Supporting Information.

\section{Summary and discussion}
Motivated by inconclusive evidence for supporting the current WHO guidelines promoting immediate ART initiation in adolescents, we have conducted an analysis comparing dynamic 
treatment initiation rules.  Our approach utilizes the theory of causal inference for DTRs.  
We extend the framework to allow the causal comparisons of both specific regimes and regimes along a continuum, 
Additionally, propose strategies to address  sparse outcomes and death, and use a composite outcome that can be used to draw causal comparisons between DTRs.

Our analysis suggests that immediate ART initiation leads to mortality benefit and higher median values of the composite outcome, 
relative to delayed ART initiation. The `never treat' regime yields significantly higher mortality than other initiation rules.

The data from IeDEA pose several challenges that we addressed within our analysis.
First, treatment initiation times are recorded on a continuous time scale.  Existing
approaches have relied primarily on discretization of the time axis to construct inverse
probability weights.  We have derived a method to construct weights that uses the continuous
time information.  Similar strategies have been employed in \cite{hu2017modeling} and \cite{johnson2005semiparametric};
see also \cite{lok2008statistical} for related work  in the context of structural nested mean models.

Second, CD4 counts are measured at irregularly spaced times.  This creates challenges
when the goal is to compare treatment regimes at a specific follow up time, as would be
the case with a randomized trial.  Moreover, even though our sample comprises those who
would be scheduled to have at least two years of follow up, some individuals discontinue
follow up prior to that time.  These features of the data lead to incomplete observation
of CD4 count at the target analysis time and to censoring of death times.
To address this issue
we have relied on a parametric model for the joint distribution of observed CD4 counts
and death times.  The CD4 submodel is flexible enough to capture important features of the
longitudinal trajectory of CD4 counts, and is used to impute missing observations at
the target follow up time.  The mortality submodel, which depends explicitly on the CD4 trajectory,
is used to impute mortality status at the target estimation time. A limitation of the imputation model is that death and CD4 may depend on HIV viral load, but availability of this variable is limited in our data and therefore not included in the model. 

The primary strength of this approach is its ability to handle a complex data
set on its own terms, without artificially aligning measurement times. Although imputation-based analyses rely on extrapolating missing outcomes, 
and both the weight model and imputation model must be correctly specified,
a potential advantage of our approach over g-computation is reduced dependence
on data extrapolation. There are
several possible extensions as well.  First, largely due to limitations related to computing,
we used a two-step approach to fit our observed-data imputation model rather than a joint
likelihood approach.  There may be
some small biases \citep{rizopoulos2012joint} introduced by using a two-step
rather than fully joint model.  Second, the imputation model may not be fully compatible
with the weighting model in the sense that we are not constructing a joint distribution
of all observed data.  Our approach emulates a setting whereby the data imputer and the
data analyst are separate:  the imputed dataset can be turned over for whatever kind of
analysis would be applied to a complete dataset.  
Empirical checks to our joint model for CD4 and mortality showed no evidence
of lack of fit to the observed data (see Supporting Information).  To make the models more flexible,
it may be possible to employ machine learning methods as in \cite{shen2017estimation}.  
Finally, developing sensitivity analyses to capture the effects of unmeasured confounding
for our model would be a worthwhile
and important contribution.

\section*{Acknowledgements}
The authors are grateful to Michael Daniels for helpful comments and to Beverly Musick 
for constructing the analysis dataset.
This work was funded by grants
R01-AI-108441, R01-CA-183854, U01-AI-069911, and P30-AI-42853 from the U.S.\ National Institutes of Health. 

 \bibliographystyle{biom} 
 \bibliography{dissertation.bib}

%
%
%



%

\label{lastpage}

\section{Supporting Information}
Additional supporting information may be found online in the Supporting Information section at the end of the article.
\backmatter

\begin{table}[H]
\centering
\caption{Summary statistics}
\label{tab:baseline}
\begin{tabular}{lcc}
\hline \hline
&\multicolumn{2}{c}{$n=1962$}\\\hline
&at $t^* = 1$ year &at $t^* = 2$ years\\\hline
ART initiated &$1286\;(65.5\%)$&$1422 \; (72.5\%)$ \\
death &$61 \;(3.1\%)$& $80 \;(4.1\%)$\\
CD4 counts per person &1.71&2.64 \\\hline
&Mean (SD) or Count (\%) & \% missing \\\hline
CD4&$343.05\; (314.78)$&21.3\%\\
WAZ&$-2.64 \;(1.83)$& 33.7\%\\
HAZ& $-2.10\; (1.48) $& 36.1\%\\
age&$12.21 \;(1.41)$ & 0\\
male&$863 \;(44.0\%)$ &0\\
CDC class &&71.6\%\\
\;mild& $200\; ( 10.2\%)$ &\\
\;moderate& $73\;(3.7\%)$ &\\
\;severe&$88\;( 4.5\%)$ &\\
\;asymptomatic &$196 \;( 10.0\%)$&  \\
person time follow up* &$3.6 \;(1.7, 6.1)$&\\\hline
\multicolumn{2}{l}{\scriptsize{$^{*}$median (1st, 3rd quartile) in years}}
\end{tabular}
\end{table}

\begin{table}[H]
\centering
\caption{Comparing effectiveness of specific regimes $q \in \{0, 200, 350, 500, \infty\}$ for  $t^* = 1$ year and $t^* = 2$ years.  $\theta_{q1} = P(X_q=0) = F_{X_q}(0)$,  $\theta_{q2} = F_{X_q}^{-1}(\half)$, $\theta_{q3} = E(X_q \cond X_q>0)$. 95\% confidence intervals are shown below the point estimates.}
\label{tab:res_dis}
 \setlength\tabcolsep{5pt}
\begin{tabular}{ccccccc}
\hline\hline
&0&200&350&500&$\infty$& $\infty$ vs. 500\\\hline
$t^*=1$ &\\
\; $\wh{\theta}_{q1}$  &.050&.018&.017& .020& .012& $-.008$\\
&(.032, .078)&(.007, .044)&(.008, .038)&(.009, .040)&(.006, .024) &$(-.015,-.001)$\\
 \; $\wh{\theta}_{q2} $ &381&292&354&375&416&41\\
 &(345, 418)&(260, 324)&(320, 387)&(350, 401)&(381, 451)&(12, 70)\\
 \; $\wh{\theta}_{q3}$ &416&326&377&401& 466 \\
 &(380, 453)&(294, 357)& (349, 406)&(373, 429)  & (435, 498) \\\hline
 $t^*=2$ &\\
 \; $\wh{\theta}_{q1}$ &.076 &.040 &.033&.036 & .023 &$-.013$\\
 & (014, .037)&(.050, .110)&(.021, .074)&(.019, .059)&(.021, .060)&$(-.023,-.004)$\\
\; $\wh{\theta}_{q2}$ &353&303&358&387&438&51\\
&(304, 402)&(262, 343)&(310, 407)&(341, 434)&(395, 481)&(14, 87)\\
 \; $\wh{\theta}_{q3}$  &394&345&388&418&484\\
 &(348, 441)&(308, 382)& (352, 423)&(381, 455)& (447, 522) \\\hline
\end{tabular}
\end{table}

%
%

\end{document}